%% file: paper.tex
\documentclass[11pt,a4paper]{article}

\usepackage{amssymb}
\usepackage{graphicx}
\usepackage{subfigure}
\usepackage{makeidx}
\usepackage{amsmath}
\usepackage{amsfonts}
\usepackage[linktocpage]{hyperref}
\usepackage[usenames,dvipsnames]{color}
\usepackage{xmpincl}
\usepackage{cite}

\usepackage{multirow}

\usepackage{parskip,url}

\def\lsim{\raise0.3ex\hbox{$<$\kern-0.75em\raise-1.1ex\hbox{$\sim$}}}
\def\gsim{\raise0.3ex\hbox{$>$\kern-0.75em\raise-1.1ex\hbox{$\sim$}}}

\newcommand*{\OriginalQuotation}{}
\let\OriginalQuotation\quotation
\renewcommand*{\quotation}{\OriginalQuotation\small\sf}
\usepackage[font=sf,textfont=small]{caption}

\begin{document}

\let\endtitlepage\relax

\input{title.tex}

\section{Introduction}

The nucleon sigma term, and the nucleon strangeness content are
phenomenologically important quantities that are not directly
accessible to experiment. They are related to $\pi - N$ and $K - N$
scattering lengths, to the quark-mass ratio $m_s/m_{ud}$ (where
$m_{ud}=(m_u+m_d)/2$), to the hadron spectrum and even to counting
rates in Higgs boson searches. They also play a key role in the direct
detection of dark matter (DM): sigma 
terms relate the effective DM - nucleon coupling to the
fundamental Lagrangian parameters that couple the DM particle to the
quarks. A precise first principle determination of these quantities is
thus very interesting. The sigma terms are defined as
\begin{subequations}
  \label{eq:sgt}
  \begin{eqnarray}
    \sigma_{\pi N} &=& \langle N(p) | m_{ud} (\bar uu+\bar dd)(0) |
                       N(p) \rangle \\
    \sigma_{\bar ss N} &=& \langle N(p) | m_s\bar ss(0) |
                           N(p) \rangle
  \end{eqnarray}
\end{subequations}

The usual way of computing $\sigma_{\pi N}$ is by using $\pi - N$
scattering data. $\sigma_{\pi N}$ cannot be directly
obtained in this way, but available data can be extrapolated to the
Cheng-Dashen point, although the machinery involved in this
determination is complicated. $\chi$PT can then be used to 
reliably extrapolate from the Cheng-Dashen point to the chiral
limit~\cite{Koch:1982pu,Gasser:1990ce}. 

Such an analysis was carried out in the
1980's~\cite{Koch:1982pu,Gasser:1990ce}, resulting in a value of
$\sigma_{\pi N} = 45\pm 8$ MeV. A later study claims to obtain a
higher 
value~\cite{Pavan:2001wz}, $\sigma_{\pi N} = 64\pm 7$ MeV which is
almost 2 standard deviations away. Also it has been pointed that the
uncertainties associated 
with these determinations are already affecting the interpretation of
(direct) dark matter search experiments~\cite{Ellis:2008hf}, although
lattice simulations are already helping to change this
situation~\cite{Giedt:2009mr}. 

Sigma-terms of other octet members are also of phenomenological
interest. Recently they have been used in the context of the hadron
resonance gas model for estimating quark-mass effects in QCD
thermodynamics calculations~\cite{Borsanyi:2010bp}.  Moreover, as we
  will see, an analysis of all octet members can be further used to
  constrain the strangeness content of the nucleon.

In this paper we compute the sigma terms of the octet
baryons $\sigma_{\pi X}, \sigma_{\bar ss X}$ for $X=N,\Lambda, \Sigma,
\Xi$. The Feynman-Hellman theorem applied to QFT relates the sigma
terms to the dependence of baryon masses with respect to quark masses
through:  
\begin{subequations}
  \begin{eqnarray}
    \sigma_{\pi X} &=& m_{ud}\frac{\partial M_X}{\partial m_{ud}} \\
    \sigma_{\bar ss X} &=& m_s \frac{\partial M_X}{\partial m_{s}}
  \end{eqnarray}
\end{subequations}
This opens the possibility of computing the octet sigma terms via the
variation of octet masses with respect to quark masses in a lattice
simulation, which is the approach that we use here. A preliminary
account of this work has been given in~\cite{Durr:2010ni}. There
are other interesting quantities directly related
to the previously mentioned sigma terms that are also useful for
phenomenology
\begin{subequations}
  \begin{eqnarray}
    \label{eq:str}
    y_X &=& \frac{2\langle X(p) | \bar ss(0) | X(p) \rangle}
    {\langle X(p) | (\bar uu+\bar dd)(0) | X(p) \rangle}  \\
    \label{eq:sgd1}
    f_{ud X} &=& \frac{m_{ud}\langle X(p) | (\bar uu+\bar dd)(0) |
      X(p) \rangle}{M_X} \\
    \label{eq:sgd2}
    f_{\bar ss X}  &=& \frac{m_s\langle X(p) | \bar ss(0) |
      X(p) \rangle}{M_X}.
  \end{eqnarray}
\end{subequations}
The first is called the strangeness content, and the other two are
sometimes referred to as ``\emph{dimensionless sigma terms}''. The
latter two
are directly related to the conversion of the fundamental 
DM-quark coupling to the effective DM-nucleon coupling.

\section{Simulation details and ensembles}

The gauge and fermionic actions,
as well as the algorithms used are described
in~\cite{Durr:2008rw,Durr:2008zz}. Here it suffices to mention that
we simulate QCD with two degenerate light quarks and a heavier strange
quark, and that we use tree level improved Wilson fermions. 

To set the lattice spacing corresponding to each value of the coupling
$\beta$ and fix the quark masses to their physical values, we use 
$M_\pi$, $M_K$, and $M_\Omega$. We extrapolate the values of $aM_\pi,
aM_K, aM_\Omega$ to the point where the ratios $aM_\pi/aM_\Omega$ and 
$aM_K/aM_\Omega$ agree with the experimental
values\footnote{Experimental inputs are corrected for isospin breaking
and electromagnetic effects according to~\cite{Durr:2008zz}.}.

\begin{table}[t!]
  \centering
  \input{stout.tex}
\caption{Parameters of our simulations.
  The errors quoted here are purely statistical. These results
  correspond to one of the 18 two-point function, time fit intervals
  that we use in our estimate of systematic uncertainties. In this
  particular analysis, the scales at $\beta=3.3, 3.57, 3.7$ are
  $a^{-1}=1616(20)$ MeV, $2425(27)$ MeV, $3142(37)$ MeV, respectively.}
  \label{tab:stout}
\end{table}

\begin{figure}
  \centering
  \includegraphics[width=12cm]{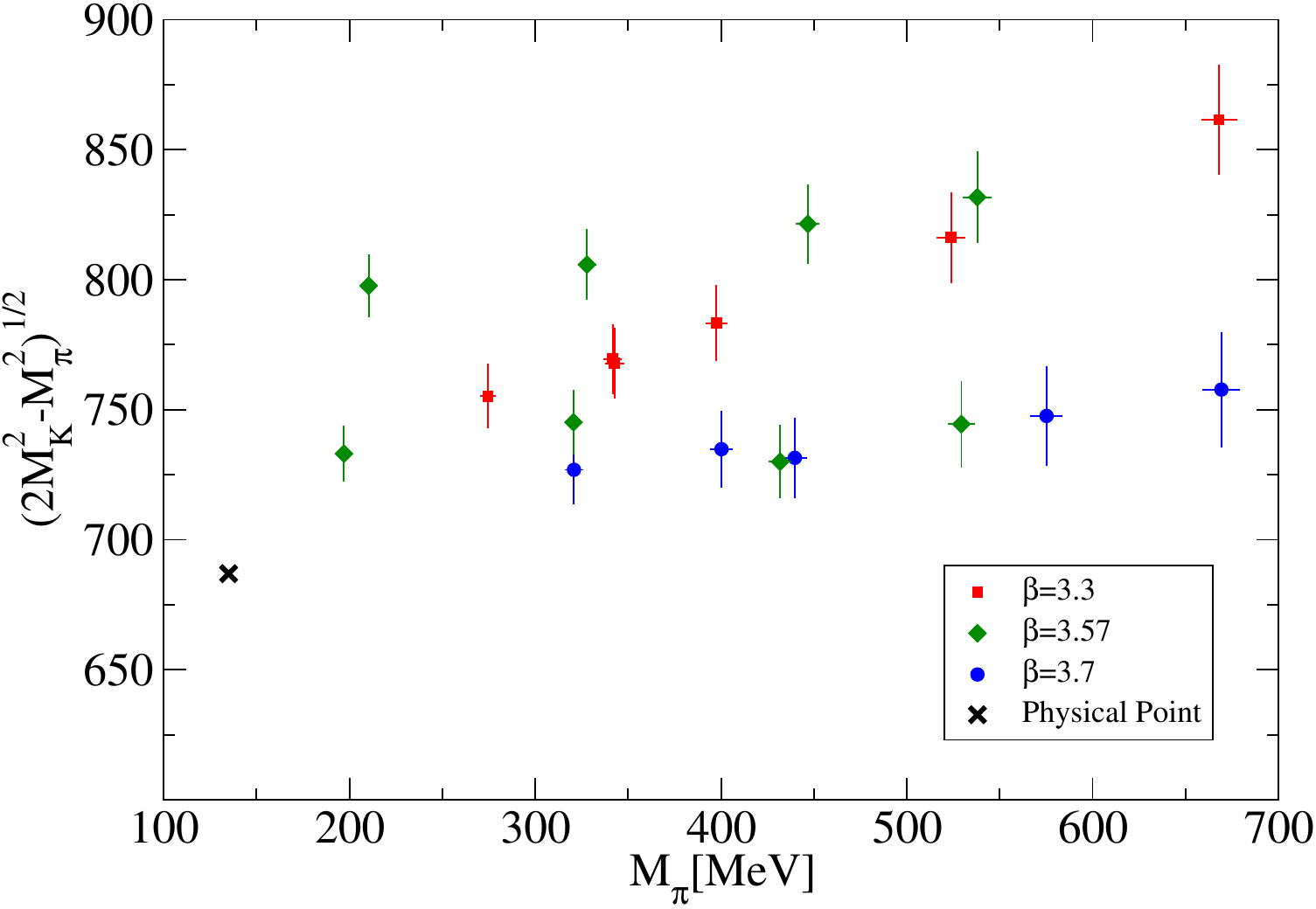}
  \caption{Overview of our simulation points in terms of $M_\pi$ and
    $\sqrt{2M_K^2 - M_\pi^2}$. The former gives a measure of the
    isospin averaged up and down quark mass while the latter determines
    the strange quark mass. The symbols refer to the three lattice
    spacings, and the physical point is marked with a cross. Error
    bars are statistical only.}
\label{fig:landscape_stout}
\end{figure}

As shown in Table~\ref{tab:stout} at $\beta=3.3$ and
$\beta=3.7$ the strange quark mass is held fixed, whereas for
$\beta=3.57$ we simulate at three different values of $m_s$ to have some
lever arm to perform the small extrapolation to $m_s^{\text{phys}}$ (see
also Fig.~{\ref{fig:landscape_stout}}). Our data sets cover a wide
range of pion masses from $M_\pi \sim 190$ MeV up to $M_\pi \sim 680$
MeV, although in this analysis we only use ensembles with $M_\pi <
550$ MeV. 

On every ensemble we measure the octet masses $M_N, M_\Lambda,
M_\Sigma$ and $M_\Xi$ with valence quark masses equal to sea quark
masses (only the unitary theory is considered). It is worth mentioning
that these ensembles have previously been used to accurately predict the
light hadron spectrum~\cite{Durr:2008zz}, including the masses of the
octet baryons.

\section{Chiral extrapolation}

As already stated, we simulate QCD for values of the light quark masses
larger than the physical values, whereas the strange quark is close to
its physical value with some lever arm to perform a small
extrapolation. Thus our data requires an extrapolation in the light
quark masses and an interpolation in $m_s$.

For this, we need to describe the quark-mass dependence of octet
baryons. We will always use the tree level $SU(3)$ chiral relation to
express the quark-mass dependence through the meson mass dependence via
\begin{eqnarray}
  \nonumber
  m_{ud} &\propto & M_\pi^2\\
  m_s &\propto & M_{\bar ss}^2 = 2M_{K}^2-M_\pi^2
  \;.
  \label{Mss}
\end{eqnarray}

To study this mass dependence and the associated model uncertainty, we
consider two very different
approaches~\cite{Durr:2008zz,Lellouch:2009fg}. First we 
consider a regular expansion in quark masses around a non 
singular point where none of the quark masses vanish. For this
purpose we define the expansion variables
\begin{eqnarray}
  \label{eq:defdeltas}
  \nonumber
  \Delta_\pi &=& \frac{M_\pi^2 - \left(
      M_\pi^{\text{cen}}\right)^2}{M_\Omega^2} \\
  \Delta_{\bar ss} &=& \frac{M_{\bar ss}^2 - 
    \left(M_{\bar ss}^{\text{phys}}\right)^2}{M_\Omega^2},
\end{eqnarray}
where $\left(M_\pi^{\text{cen}}\right)^2 =
\frac{1}{2}\left[\left(M_\pi^{\text{phys}}\right)^2 +
  \left(M_\pi^{\text{cut}}\right)^2\right]$, and $M_\pi^{\text{cut}}$ denotes
the mass of the heaviest pion included in the fit. 

Note that changing the expansion point in polynomial-like 
formulae is simply a reshuffling of the coefficients. When
interested in checking the $SU(3)$ symmetric line $m_{ud} = m_s$ it is
far more convenient to define our expansion variables as 
\begin{eqnarray}
  \label{eq:defgammas}
  \nonumber
  \Delta_\pi &=& \left(\frac{M_\pi}{M_\Omega}\right)^2 \\
  \Delta_{\bar ss} &=& \left(\frac{M_{\bar ss}}{M_\Omega}\right)^2,
\end{eqnarray}
where the $SU(3)$ symmetric line is characterized by $\Delta_\pi =
\Delta_{\bar ss}$. Because the extrapolation in $m_s$ is small, a
linear term in $\Delta_{\bar ss}$ suffices 

Second we consider a singular expansion around the $SU(3)$ chiral
point $m_{ud} = m_s = 0$, which allows for a much more constrained
expansion. Low-energy processes in QCD have been
extensively  
studied in the framework of chiral perturbation theory. The
seminal work~\cite{Gasser:1983yg} and its success in
explaining meson observables, have pushed the study of baryons within
the same framework. However baryon $\chi$PT is far more involved. The
main difficulty is that  
baryon masses $M_B$ are not small in comparison with the scale of chiral
symmetry breaking ($\Lambda_\chi \sim 1$ GeV), and Weinberg's
power counting theorem~\cite{Weinberg:1978kz} fails: higher-order loop
corrections contribute with powers of $M_B/\Lambda_\chi$ and are no
longer small. This results in a very slow (practical) convergence of 
the series~\cite{Gasser:1987rb}. This convergence is partially
improved by treating the baryons as heavy degrees of freedom, in what
is known as the heavy baryon chiral perturbation
theory~\cite{Jenkins:1990jv}. Although it is reasonable for dealing with
the mass dependence of baryons with light quarks, experience seems to
show that observables that depend on the strange quark are not well
described in this framework (see for
example~\cite{Borasoy:1996bx, Donoghue:1998bs, Borasoy:1998uu}). A
number of possibilities have been proposed to improve the convergence
and include the strange quark in the analysis. The various flavors of
cutoff B$\chi$PT (see~\cite{Young:2002ib} and references
therein), or covariant B$\chi$PT~\cite{Dorati:2007bk} are some
examples. In this work we explore this last possibility. The
interested reader can consult the review~\cite{Bernard:2007zu} and
references therein.

\subsection{Regular expansions}

We have several possibilities if we decide to expand around a
regular point. We can treat any
octet member $X=N,\Lambda,\Sigma,\Xi$ as independent and use a Taylor
expansion. Using the variables defined in Eq.~(\ref{eq:defdeltas}) our
results are well described by the ansatz 
\begin{equation}
  M_X = M_0^X + \alpha_1^X \Delta_\pi + \alpha_2^X \Delta_\pi^2 + 
  \beta^X \Delta_{\bar ss}.
\end{equation}
One can also use a Pad\'e like functional form
\begin{equation}
  M_X = \frac{M_0^X}{1 - \alpha_1^X \Delta_\pi - \alpha_2^X \Delta_\pi^2 -
  \beta^X \Delta_{\bar ss} }.
\end{equation}
In these expansions, $M_0^X, \alpha_{1,2}^X$ and $\beta^X$ are the 16
fitting parameters. 

In principle all octet masses should be degenerate along the $SU(3)$
symmetric line $m_{ud} = m_s$. We can choose to impose this constraint
in our functional form. In fact our data is well fitted by the
following $SU(3)$ symmetric regular expansion
\begin{equation}
  M_X = M_0 + \alpha_1^X \Delta_\pi + \alpha_2^X \Delta_\pi^2 + 
  \left(C_1 - \alpha_1^X\right) \Delta_{\bar ss} + 
  \left(C_2 - \alpha_2^X\right) \Delta_\pi\Delta_{\bar ss},
\end{equation}
or the corresponding Pad\'e like ansatz
\begin{equation}
  M_X = \frac{M_0}{1 - \alpha_1^X \Delta_\pi - \alpha_2^X \Delta_\pi^2 - 
  \left(C_1 - \alpha_1^X\right) \Delta_{\bar ss} - 
  \left(C_2 - \alpha_2^X\right) \Delta_\pi\Delta_{\bar ss}}.
\end{equation}

Now the fitting parameters have been reduced to 11: $M_0,
\alpha_{1,2}^X$ and the coefficients $C_1$ and $C_2$ that give the
dependence of octet masses along the $SU(3)$ symmetric line $m_{ud} =
m_s$. 

Note that Taylor-like and Pad\'e-like ansatze differ in higher-order
terms in $\Delta_\pi$ and $\Delta_{\bar ss}$. Thus the
difference in physical results obtained by using these two 
functional forms measures higher-order contributions to the mass
expansion.

\subsection{Covariant B$\chi$PT}
\label{sect:cbchipt}

Details of the quark-mass dependence of the octet members in $SU(3)$
baryon $\chi$PT can be found in Appendix~\ref{ch:cpt}. Here
we run the index $X = N, \Lambda, \Sigma, \Xi$
over octet members, and index $\alpha = \pi, K, \eta$ over the
mesons. 

Octet masses to NLO involve the nonlinear function
\begin{equation}
  h(x) = -\frac{x^3}{4\pi^2} \left\{
    \sqrt{1-\left(\frac{x}{2}\right)^2}\arccos{\frac{x}{2}}
    + \frac{x}{2}\log x
  \right\}.
\end{equation}

The masses of the octet baryons are given by
\begin{equation}
  \label{eq:chpt}
  M_X = M_0 - 4c_{\pi X} M_\pi^2 - 4c_{\bar ss X}M_{\bar ss}^2 + 
  \sum_{\alpha=\pi,K,\eta} \frac{g_{\alpha X}}{F_\alpha^2} M_0^3
  h\left(\frac{M_\alpha}{M_0}\right) + 4d_\pi M_\pi^4 + 4d_{\bar ss}
  M_{\bar s s}^4,
\end{equation}
where $c_{\pi X}, c_{\bar ss X}$ are functions of the three low-energy
constants (LEC), $b_0,b_d,b_f$, and the $g_{\alpha X}$ can be written
in terms of the axial coupling $g_A$ and the ratio of couplings $\xi$
(as shown in Tabs.~\ref{tab:gxa} and~\ref{tab:lo}). Finally $d_\pi,
d_{\bar ss}$ parametrize higher-order corrections.  
\begin{table}[h!]
  \centering
  \input{gxa.tex}
  \caption{Meson-loop couplings as a function of the $\pi N$ coupling
    $g_A$ and the quantity $\xi$. In the chiral limit $g_A = D+F$ and
    $\xi=F/D$, as described in Appendix~\ref{ch:cpt}.}
  \label{tab:gxa}
\end{table}

To the order at which we are working only two pseudo-Goldstone boson
masses are linearly independent, being related through the
Gell-Mann-Okubo relation $3M_\eta^2 = 4M_K^2 - M_\pi^2 $. 
\begin{table}[h!]
  \centering
  \input{lo.tex}
  \caption{Leading order octet quark-mass dependence, $c_{\pi X}$ and
    $c_{\bar ss X}$, 
    as a function of the LEC $b_0,b_D,b_F$ (see Appendix~\ref{ch:cpt}
    for more details).} 
  \label{tab:lo}
\end{table}

When fitting lattice data to the chiral formula, it is important not
to break the $SU(3)$ symmetry built into this expression. In that
sense it is important that all octet masses become degenerate when
$M_{\bar ss} = M_\pi$. This rules out the possibility of fixing the
meson decay constants $F_{\pi,K,\eta}$ to their physical values. Here
we choose to fix all of them to the common value
$F_\pi^{\text{phys}} = 92.2$ MeV. This is correct to the order at
which we are working. Note that we have repeated the fits including
the NLO terms for the decay constants and obtained results compatible
within statistical errors. We have also confirmed that the quality of
fits decreases if a non $SU(3)$ symmetric ansatz is imposed by fixing
each meson decay constant to its physical value.

The axial coupling $g_A$ is well known from phenomenology.
The most precise value at the physical point is
$g_A =1.2695(29)$~\cite{Nakamura:2010zzi}, and is expected to be close to the 
value in the chiral limit $D+F$. The ratio of couplings 
$\xi$ is not well 
determined experimentally, but there are two preferred
phenomenological scenarios: $\xi = 2/3$ and $\xi \sim 0.5$~\cite{Cheng:1997tt}. 

\begin{figure}[h!]
  \centering
  \subfigure[Fit in the $M_\pi^2,M_X^2$ plane. Data have been 
  corrected to the physical value of
  $M_{\bar ss}$ for a better visualisation
  ]{\includegraphics[width=7.3cm]{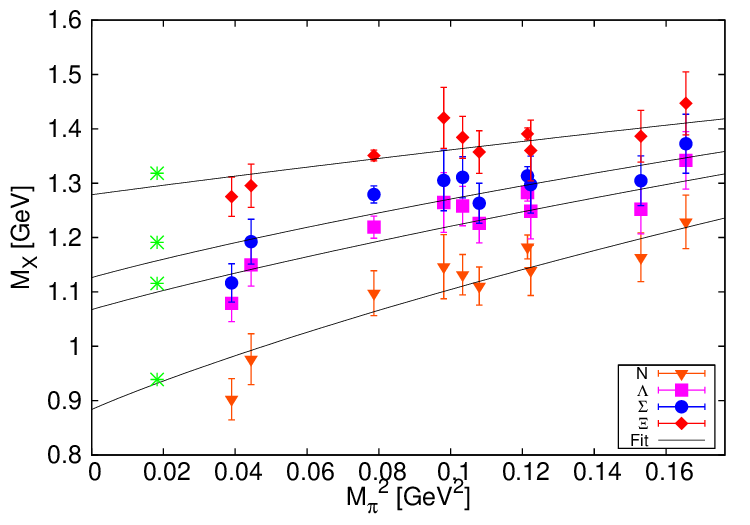}}\quad 
  \subfigure[Fit in the $M_{\bar ss}^2,M_X^2$ plane. Data have been 
  corrected to the physical value of
  $M_{\pi}$ for a better
  visualisation]{\includegraphics[width=7.3cm]{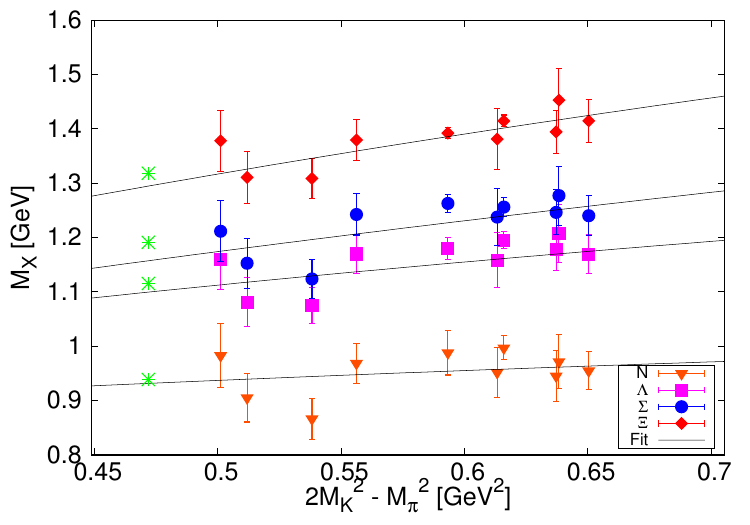}}\\ 
  \caption{Example of a $\chi$PT fit corresponding to a particular
    choice of fitting interval for the correlators. In this particular
    case we choose to fit $g_A$ and $\xi$, and not to include
    higher-order terms. We have only used data with $M_\pi < 410$
    MeV. The 
    correlated $\chi^2 = 38.5$ for 34 degrees of freedom,
    yields a fit quality of $\sim 0.27$.} 
  \label{fig:chpt}
\end{figure}

It makes sense to try our $\chi$PT fits both fixing $(g_A,\xi) =
(1.2695,2/3)$ or allowing them to be free. Regarding the higher-order
contributions given by $d_\pi, d_{\bar ss}$, we can also choose to
either include them in the fit, or not. In terms of how well our data
are described (i.e. fit quality) the four possibilities are equally
reasonable options. In Fig.~\ref{fig:chpt}, we show one possible
$\chi$PT fit, where we have fitted $g_A, \xi$ and we have not included
the higher-order terms in the fit.

\section{Cutoff, finite volume effects and excited-state contributions }
\label{sc:fv}

We combine our chiral extrapolation with the continuum
extrapolation. It has been demonstrated in different
contexts~\cite{Capitani:2006ni,Hoffmann:2007nm,Durr:2008zz,Durr:2008rw}
that a smeared clover action is very close to be nonperturbatively
$\mathcal O(a)$-improved with cutoff effects being $\mathcal
O(a^2)$. Nevertheless we cannot rule out the possibility that our
results have linear discretization effects.

Guided by our experience~\cite{Durr:2008zz} in determining light
hadron masses we parametrize cutoff effects with the substitution
\begin{equation}
  M_X \longrightarrow M_X(1+c_Xa^p) 
\end{equation}
with $p=1,2$. The cutoff effects are small
enough in our data that they make it difficult to distinguish $a$ and
$a^2$. In fact their coefficients are, for the most part,
compatible with zero within statistical errors. Thus 
we will also include fits without any cutoff corrections altogether.

Stable hadron masses have leading finite-volume effects $\propto
e^{-M_\pi L}$~\cite{Luscher:1985dn}. In all of our ensembles the bound
$M_\pi L \gsim 4$ is maintained, implying negligible finite-volume
effects. The detailed analysis of~\cite{Durr:2008zz}, with 
additional ensembles in smaller volumes, shows that finite-volume
corrections are below the statistical accuracy of our data. Thus we
do not add finite-volume corrections to our analysis here.

To estimate the possible contamination by excited states in the
extraction of the masses from the correlators we use 18 fitting
time intervals: $t_{\text{min}}/a = 5$ or 6 for $\beta=3.3$,
$t_{\text{min}}/a = 7, 8$ or 9 for $\beta=3.57$ and 
$t_{\text{min}}/a = 10, 11$ or 12 for $\beta=3.7$ \cite{Durr:2008zz}.

\section{Determination of the uncertainties}

Our complete analysis includes a total of 8 formulae to extrapolate to
the physical mass point: Four of them are regular expansions either
imposing or not the $SU(3)$ flavor constraint, and either in the
Taylor-like form or Pad\'e-like form. The other four functional forms are
derived from $SU(3)$ covariant baryon $\chi$PT, where we can choose to
parametrize or not higher-order contributions, and to fit $g_A$ and $\xi$
or fix $g_A$ to its physical value
$g_A=1.2695(29)$~\cite{Nakamura:2010zzi} and $\xi$ to a reasonable
phenomenological value\footnote{We have checked that using
  $\xi = 0.5$ (the other common phenomenological scenario) leads to
  very similar results.}  $\xi=2/3$~\cite{Cheng:1997tt}. 

To have more control over the uncertainties associated with
higher-order terms in the mass extrapolation, we
impose two different pion mass cuts: $M_\pi < 410$ MeV and
$M_\pi < 550$ MeV. 

As explained in Sec.~\ref{sc:fv} we consider three possibilities to
parametrize cutoff effects which are compatible with our data: either
assume that they are absent, or parametrize them as $\mathcal O(a)$ or
$\mathcal O(a^2)$. Finite-volume corrections are not considered
because they are below our statistical
accuracy. Finally we repeat the full analysis with 18 different
time-fitting intervals for the correlators to estimate excited state 
contributions.  

\begin{figure}
  \centering
  \subfigure[Final distribution of values for $\sigma_{\pi N}$ and the
  contributions of different pion mass cuts.]{\label{fig:mpi}\includegraphics[width=7.3cm]{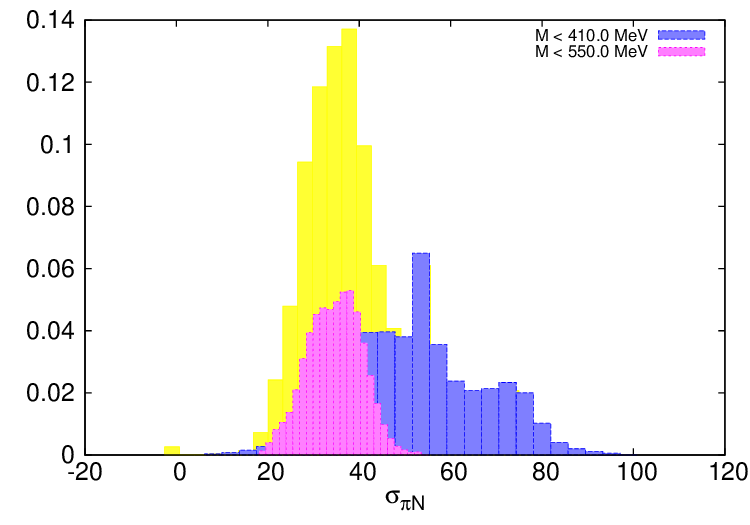}}\quad 
  \subfigure[Final distribution of values for $\sigma_{\pi N}$ and the
  contributions of different functional forms.]{\label{fig:mth}\includegraphics[width=7.3cm]{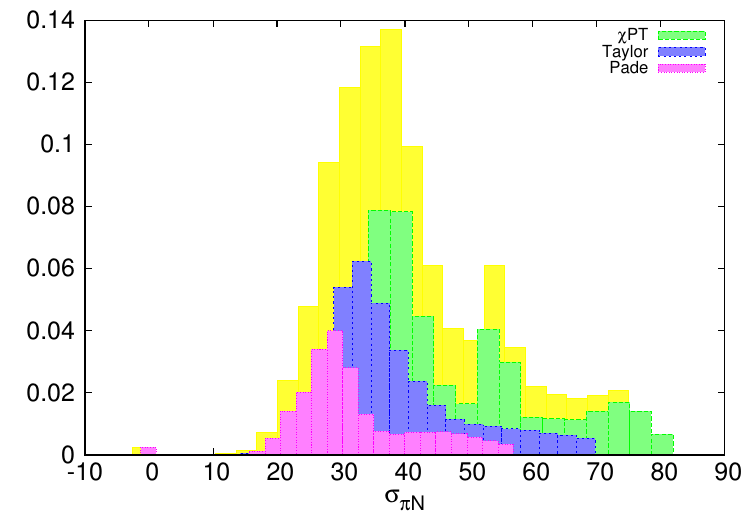}}\\
  \caption{Distribution of values for $\sigma_{\pi N}$ in background
    (yellow), and how different pion mass cuts (left) and functional
    forms (right) shape it.}
  \label{fig:dist}
\end{figure}

This strategy leads to $8\times 2\times 3\times 18=864$ different
procedures to estimate the physical value of each of the sigma terms of the
octet members. The data are fitted (taking correlations into account)
using all of the previously explained procedures. Each of the 864
results is weighted with the fit quality
\footnote{The confidence of fit is defined as
  $Q=\int_{\chi^2}^\infty dz \; P(z,d)$ where
  $P(z,d)=z^{d/2-1}e^{-z/2}/[2^{d/2}\Gamma(d/2)]$ is the
  probability distribution function to obtain $\chi^2=z$
  in a fit with $d$ degrees of freedom.}
  ($p$-value or $Q$-value) to produce a
distribution of values. The median (typical result of our analysis) is
taken as our final value. The
$16^{\underline{\text{th}}}$/$84^{\underline{\text{th}}}$ percentiles
(i.e.\ the values which, in a Gaussian distribution, correspond to
$\pm1\sigma$ deviations from the median)
yield the systematic uncertainty of our computation (see
Fig.~\ref{fig:dist}). This systematic 
uncertainty measures how different parametrizations for mass 
extrapolation, cutoff effects and excited-state
contributions affect the final result. 
The statistical uncertainty is determined by bootstraping the whole
procedure 2000 times, and computing the variance of the medians. 

It is important to note here three points: 
\begin{itemize}
\item  Most of the previously
  mentioned fits are ``good.'' Our average fit quality is $0.34$, and
  this does not vary significantly over our different procedures. For
  example fits with $M_\pi < 410$ MeV have an average fit quality of
  $0.38$ whereas for $M_\pi < 550$ MeV it is $0.29$ (see
  Fig.~\ref{fig:mpi}). For covariant B$\chi$PT, Taylor 
  and Pad\'e functional forms the p-values are respectively $0.30,
  0.47$ and 
  $0.29$ (see Fig.~\ref{fig:mth}). The fit quality is almost
  insensitive to how we choose to 
  parametrise the cutoff effects, or what time intervals we use to fit
  correlators. Although there is some variation, we
  cannot rule out any of our fits based on how well our data
  are described.
\item Adding more variations does not increase the systematic
  uncertainty. For example adding a pion mass cut of $M_\pi < 680$ MeV
  (i.e. including all our ensembles in the fit), the covariant B$\chi$PT
  functional form gives fit qualities of $2\times 10^{-6}$. Even if
  they are included in our analysis, these fits do not contribute
  to our distribution of values or to our final results and
  estimates of systematic uncertainties.
\item On the other hand, \emph{eliminating} some of the analysis 
  typically results in a compatible final value but with a smaller
  systematic uncertainty. For example had we chosen to perform
  our analysis only with $M_\pi < 550$ MeV the systematic uncertainty
  of $\sigma_{\pi N}$ would have dropped by almost $50\%$. (see
  Fig.~\ref{fig:mpi}). 
\end{itemize}

\section{Results}

Following the method described in the previous section, we arrive at
the results quoted in Table~\ref{tab:res}.
\begin{table}[t!]
  \centering
  \input{res.tex}

  \caption{Final results for all octet members. The first two columns
    give the baryon octet sigma terms (Eq.~(\ref{eq:sgt})), the third
    one the strangeness content of the octet member (Eq.~(\ref{eq:str})),
    and the last two the dimensionless sigma terms (Eqs.~(\ref{eq:sgd1}),
    (\ref{eq:sgd2})). The first error
    is statistical, the second one systematic. }
  \label{tab:res}
\end{table}

For the nucleon sigma term our result agrees with the ``canonical'' 
determination coming from $\pi-N$ scattering
data~\cite{Gasser:1990ce} of $\sigma_{\pi N} = 45\pm 8$ MeV, but with
a larger uncertainty, and disfavors the larger value $\sigma_{\pi
N} = 64\pm 7$ MeV from~\cite{Pavan:2001wz}. It also agrees with recent
estimates that use lattice data~\cite{Borasoy:1996bx, Leinweber:2003dg,
  Procura:2003ig, Procura:2006bj, Ohki:2008ff, Young:2009zb,
  MartinCamalich:2010fp,Roger:LAT11} though these assume a
particular functional form for 
the light quark-mass dependence while we consider a whole range of
forms. Another study considers a variety of chiral forms, but with
pion masses only down to 300~MeV~\cite{WalkerLoud:2008bp}, yielding values
in the range 42 to 84 MeV, with a statistical accuracy of about 15
MeV. Our value is less consistent with the determinations
of~\cite{Alexandrou:2008tn, Ishikawa:2009vc} that are close to the
higher value  (of~\cite{Pavan:2001wz}) for the sigma term.

We believe that the agreement within error of our results obtained
from 
extrapolations using very different functional forms enhances the
credibility of our final results.

In Table~\ref{tab:err} we give the contribution of each source of
systematic uncertainty to the final error. Although we show the
results for $\sigma_{\pi N}$, the conclusion that the chiral
extrapolation dominates the systematic uncertainty of our computation
is generally true. 
\begin{table}[h!]
  \centering
  \input{piN.tex}
  \caption{Contribution to the total uncertainty of the different
    sources of systematic error.}
\label{tab:err}
\end{table}

To reduce this source of systematic uncertainty we would need to add
ensembles at lower pion masses. From the value of the
uncertainties of $\sigma_{\pi N}$ at $M_\pi = 200$ MeV, we can roughly
guess the gain that would be obtained by having equally precise data at the physical
point. Such data would imply a reduction of around $50\%$ in the
systematic error. On the other hand the situation for both
$\sigma_{\bar ss N}$ and $y_{N}$ is less clear. Our data set has a
much smaller range of strange quark masses than light quark
masses. If we add the fact that the contribution of the strange quark
to the nucleon mass seems to be small ``per se'', it is natural that
the final results for both $\sigma_{\bar ss N}$ and $y_N$ to show large
statistical and systematic uncertainties. Only by adding simulation
points for a wider range of strange quark masses one can increase the
chances of getting a precise determination of these quantities without
making an additional assumption which relates $m_s$ to $m_{ud}$
dependence as in $SU(3)$ baryon $\chi$PT. 

Regarding the other octet members, the values of the sigma terms
agrees with other determinations based on lattice data and a
low-energy effective field theory approach~\cite{Young:2009zb, 
  MartinCamalich:2010fp}.  

\subsection{$\chi$PT consistency check}

Our objective in this paper is to determine the sigma-terms and
strangeness contents of the octet baryons at the physical point. To this
end, we generated ensembles close to the physical value of the strange
quark mass and as close as possible to the physical light quark mass,
with a large enough lever arm in the light sector to perform a
credible extrapolation to the physical point. 

Such a set of ensembles, however, is not ideally suited to determine
the $SU(3)$ $\chi$PT LECs, which are defined in the chiral limit. That
would require simulations with a strange quark mass approaching the
chiral limit. This means that the LECs extracted from our fits have
uncertainties which we cannot properly control. Nevertheless, it is
worth noting that the values obtained are roughly consistent with
general expectations from phenomenology \cite{Nakamura:2010zzi} and other
lattice studies~\cite{Young:2009zb,
  MartinCamalich:2010fp} (see Table~\ref{tab:lec}). This further
enhances the credibility of the results obtained at the physical
point.

\begin{table}[h!]
  \centering
  \input{lec.tex}
  \caption{Values of the LEC's obtained from our $\chi$PT fits. Total
    errors (statistics and systematics included) are computed in the
    same way as for the sigma terms. For the reasons noted in the
    text, we do not consider these errors to be reliable.}
\label{tab:lec}
\end{table}

\section{Conclusions}

We computed the sigma terms of light octet baryons by studying their mass
dependence on quark masses through the Feynman-Hellman
theorem. We also obtained the strangeness
content and dimensionless sigma terms for all octet members. We
estimated the systematic uncertainties associated with the choice of
parametrization for this mass dependence by considering two very
different expansion schemes. We considered regular
expansions of octet masses, and a singular three-flavor, low-energy
effective field theory 
approach. Within each of the approaches we estimated the impact of 
higher order contributions. Moreover we varied the range of fitted
quark masses, cutoff parametrizations and excited-state
contributions. 

The most interesting quantities are the ones associated with
the nucleon. Our final value for $\sigma_{\pi N} = 39(4)(^{+18}_{-7})$
favors the ``low scenario'' with $\sigma_{\pi N} = 45\pm 8$
MeV of~\cite{Koch:1982pu,Gasser:1990ce}, but the size of our
uncertainties does not allow us to exclude the higher value
of~\cite{Pavan:2001wz}. Regarding the strangeness content we obtain a
value with 
large statistical and systematic uncertainties $y_N =
0.20(7)(^{+13}_{-17})$. The dimensionless sigma terms $f_{ud N} =
0.042(5)(^{+21}_{-8})$ and $f_{\bar ss N} = 0.072(29)(^{+60}_{-50})$
are the quantities of interest for direct DM searches. 

This work shows the difficulties associated with obtaining precise
values for these form factors in a
model independent way.  Even with
pion masses as low as $190$~MeV the mass extrapolation dominates the
systematic uncertainties of our final results. This explicitly shows
the risks of using lattice data and only one approach to perform the
mass extrapolation. A proper estimation of systematic uncertainties
should include the uncertainty associated with the model that is used,
and, at least in our case, this is one of the main sources of
systematic uncertainty. 

For $\sigma_{\pi N}$ the only way to
increase the precision of our computation without assuming a model
would be to generate data at lower quark masses. Replacing the 
extrapolation by an interpolation drastically reduces the uncertainty
in the final result, as we have explicitly seen by considering the values
of the sigma terms at $M_\pi = 200$~MeV.

For the case of the strangeness content of the nucleon and $\sigma_{\bar
  ss N}$, the situation is even more delicate. First, $N_f = 2+1$
lattice ensembles 
usually bracket the physical value of $m_s$ with a small lever
arm. This is more than sufficient to determine physical observables,
but it is not 
the ideal setup to estimate how physical quantities change with
$m_s$. Second, the contribution of the strange quark to the nucleon
mass is small ``per se''. In principle the first problem can
be solved by
measuring the nucleon mass for a larger range of strange quark
masses. The second problem is solved if we restrict our analysis to
three 
flavor B$\chi$PT. This low-energy effective field theory approach
relates the $m_{ud}$ dependence of octet members to its $m_s$
dependence, and thus constrains $y_N$. 
The value of the nucleon strangeness content obtained from the covariant 
B$\chi$PT analysis alone has a small relative error $y_{N} =
0.276(77)(^{+90}_{-62})$, illustrating the success of the
idea. However, since we do not have a large enough range of strange
quark masses to establish the validity of $SU(3)$ B$\chi$PT, this
result must be considered model dependent. 
Note also that a direct calculation of the corresponding disconnected
contribution suggests a very small $y_N$ value, barely compatible with
the B$\chi$PT result \cite{Andreas:LAT11}.

\section*{Acknowledgments}

Computations were performed using HPC resources from FZ J\"ulich and
from GENCI-[IDRIS/ CCRT] (Grant No. 52275) and clusters at Wuppertal and
CPT. This work is supported in part by EU Grants No. I3HP, 
FP7/2007-2013/ERC No. 208740, No. MRTN-CT-2006-035482 (FLAVIAnet), DFG Grant
No. FO 502/2, SFB-TR 55, by CNRS Grant GDR No. 2921 and PICS No. 4707. 

The authors also thank Martin Jung for pointing out an error in
the normalization of the dimensionless sigma terms.

\appendix

\section*{Appendix}

\section{Sketch of three-flavor covariant baryon $\chi$PT}
\label{ch:cpt}

In this appendix, we give a brief outline of the derivation of our
chiral fit form (\ref{eq:chpt}). Working to ${\cal O}(p^3)$, we use
the standard octet baryon Lagrangians
\cite{Oller:2006yh,Frink:2006hx}
\begin{eqnarray}
{\cal L}_{MB}^{(1)}&=&\text{Tr}\Big(
\text{i}{\overline B}\gamma^\mu D_\mu B-M_0{\overline B} B+
\frac{D}{2}{\overline B}\gamma^\mu\gamma_5\{u_\mu,B\}+
\frac{F}{2}{\overline B}\gamma^\mu\gamma_5[u_\mu,B]
\Big)
\label{p1}
\\
{\cal L}_{MB}^{(2)}&=&\text{Tr}\Big(
b_D{\overline B}\{\chi_+,B\}+b_F{\overline B}[\chi_+,B]+\dots
\Big)+
b_0\text{Tr}\Big({\overline B}B\Big)\text{Tr}\Big(\chi_+\Big)+\ldots
\end{eqnarray}
with the chiral tensors
\begin{eqnarray}
D_\mu B&=&\partial_\mu B+[\Gamma_\mu,B]
\\
\Gamma_\mu&=&\frac{1}{2}(u^\dagger(\partial_\mu+\ldots)u+u(\partial_\mu+\ldots)u^\dagger)
\\
u_\mu&=&\text{i}(u^\dagger(\partial_\mu+\ldots)u-u(\partial_\mu+\ldots)u^\dagger)
\\
\chi_\pm&=&u^\dagger\chi u^\dagger\pm u\chi^\dagger u
\\
\chi&=&2B_0(M+\ldots),
\end{eqnarray}
where $M$ denotes the (diagonal) quark-mass matrix.  To the order we
are working, there is no contribution from the $\mathcal O(p^3)$
Lagrangian, ${\cal L}_{MB}^{(3)}$, to 
the quark-mass dependence of the octet baryon masses.
The coupling constants $D,F$ of Eq.~(\ref{p1}) are related to the
actual couplings of the nucleon in the chiral limit through
$D+F=\lim_{u,d,s\to0}g_A$ and $F/D=\lim_{u,d,s\to0}\xi$, whereas $M_0$
corresponds to the chiral limit value of the mass of the baryon octet.
The field $u(x)=\sqrt{U(x)}$ describes a nonlinear matrix
representation of the (quasi-) Goldstone boson fields and $B$ is the
matrix-valued interpolating field for a spin 1/2 octet baryon.  The
next-to-leading order low-energy constants $b_D,b_F,b_0$ govern the
leading quark-mass contributions to the mass of a spin 1/2 octet
baryon, with ${\cal M}$ denoting the quark-mass matrix for three light
flavors $u,d$ and $s$.  Furthermore, we note that the parameter $B_0$ is a
measure of the size of the chiral condensate
$\langle0|\bar{q}q|0\rangle$ (in the chiral limit)~\cite{Gasser:1984ux}
\begin{equation}
B_0\equiv-\,\frac{\langle0|\bar{q}q|0\rangle_{m_q\to0}}{F_0^2}
\end{equation}
where the low-energy constant $F_0$ is identified with the value of the octet
Goldstone boson decay constant in the chiral limit.
Finally, for the calculation at hand one needs to know the chiral meson
Lagrangian up to ${\cal O}(p^2)$ \cite{Gasser:1984ux}
\begin{eqnarray}
{\cal L}_{M}^{(2)}&=&\frac{F_0^2}{4}
\text{Tr}\{(\partial_\mu+\ldots) U^\dagger(\partial^\mu+\ldots) U+\chi^\dagger U+\chi U^\dagger\}.
\end{eqnarray}

Generalizing the SU(2) calculation of \cite{Procura:2003ig} along the
lines of \cite{MartinCamalich:2010fp}, we obtain the leading one-loop
result for the mass of a nucleon to ${\cal O}(p^3)$
\begin{equation}
\label{m3} 
\begin{split}
M_N&=M_0-4(b_0+(b_D+b_F)/2)\bar{M}_\pi^2-2(b_0+b_D-b_F)\bar{M}_{\bar{s}s}^2
\\
&-\sum_{\alpha=\pi,K,\eta}
\frac{{g}_{\alpha N}\,\bar{M}_\alpha^3}{4\pi^2F_\alpha^2}
\Big\{\sqrt{1-\frac{\bar{M}_\alpha^2}{4M_0^2}}\arccos{\frac{\bar{M}_\alpha}{2M_0}}+
\frac{\bar{M}_\alpha}{4M_0}\log\frac{\bar{M}_\alpha^2}{M_0^2}\Big\}
+{\cal O}(p^4).
\end{split}
\end{equation}

\begin{table}[tb]
\centering
\begin{tabular}{|c|cccc|}
\hline
$g_{\alpha\!B}$ & $N$ & $\Lambda$ & $\Sigma$ & $\Xi$ \\
\hline
$\pi$ & $\frac{3}{4}(D\!+\!F)^2$ & $D^2$ &
        $\frac{1}{3}(D^2\!+\!6F^2)$ & $\frac{3}{4}(D\!-\!F)^2$\\
$K$   & $\frac{1}{6}(5D^2\!-\!6DF\!+\!9F^2)$ & $\frac{1}{3}(D^2\!+\!9F^2)$ &
        $(D^2\!+\!F^2)$ & $\frac{1}{6}(5D^2\!+\!6DF\!+\!9F^2)$ \\
$\eta$ & $\frac{1}{12}(D\!-\!3F)^2$ & $\frac{1}{3}D^2$ &
        $\frac{1}{3}D^2$ & $\frac{1}{12}(D\!+\!3F)^2$\\
\hline
\end{tabular}
\caption{Summary of meson-loop couplings $g_{\alpha B}$ in terms of the
low-energy constants $D,F$.}
\label{tab:gmb}
\end{table}

To the chiral order we are working here, the meson-loop couplings
${g}_{\pi N},{g}_{KN},{g}_{\eta N}$ can be expressed
in terms of the two SU(3) parameters $F$ and $D$, as given in
Table~\ref{tab:gmb}.  For generality, we have expressed the Goldstone
boson decay constant $F_0$ entering the loop contributions via three
different symbols $F_\alpha$ to account for the individual
contribution of the pion, kaon and eta-cloud of the nucleon.

For a fit to lattice QCD data, it is convenient to slightly rewrite
(\ref{m3}) without changing the expression at the order at which we
are working: 
\begin{enumerate}
\item
We express the GMOR-masses for $\bar{M}_K,\bar{M}_\eta$ in (\ref{m3}) as a function
of the GMOR-mass of the pion and of the mass-parameter $M_{\bar{s}s}$ introduced in
(\ref{Mss}):
\begin{eqnarray}
\bar{M}_K^2&\to&\frac{1}{2}(\bar{M}_\pi^2+M_{\bar{s}s}^2)
\\
\bar{M}_\eta^2&\to&\frac{1}{3}(\bar{M}_\pi^2+2 M_{\bar{s}s}^2)
\end{eqnarray}
which amounts to a change of variables.
\item
We identify the GMOR-mass $\bar{M}_\pi$ of (\ref{Mss}) with the
corresponding lowest-lying $0^-$-boson mass $M_\pi$ in each
simulation.  Possible deviations in the quark-mass dependence from the 
linear GMOR-behavior predicted in (\ref{Mss}) can only affect ${\cal
  O}(p^4)$ corrections (\ref{m3}).
\item 
The Goldstone boson decay constants $F_\alpha$ in (\ref{m3})
are identified with the decay constant $F_\pi^\text{phys}$. Other assignments
compatible with SU(3) symmetry might be chosen as discussed in
Sec.~\ref{sect:cbchipt}.
\item
With the same reasoning as in the previous item, at chiral order
${\cal O}(p^3)$ we can identify the chiral limit couplings
$\tilde{g}_{\pi N},\tilde{g}_{KN},\tilde{g}_{\eta N}$ with the
physical couplings 
$g_{\pi N},g_{KN},g_{\eta N}$, resulting in Table~\ref{tab:gxa}.
The unknown ratio $\xi$ of 
couplings is either treated as an external input to our fits or is
left as a free fit parameter.
\end{enumerate}

With a very similar derivation for the case of the octet hyperons, we
finally arrive at our fit function (\ref{eq:chpt}).

\bibliography{/home/alberto/latex/math,/home/alberto/latex/campos,/home/alberto/latex/fisica,/home/alberto/latex/computing}

\end{document}

%% file: title.tex
\begin{titlepage}

\textbf{\LARGE{Sigma term and strangeness content of octet baryons}}

\vspace*{0.25cm}

\begin{center}
S.\,D\"urr$^{a,b}$, Z.\,Fodor$^{a,b,c}$, T.\,Hemmert$^{d}$,
C.\,Hoelbling$^{a}$, J. Frison$^{e}$, S.D.\,Katz$^{a,c}$,
S.\,Krieg$^{a,b}$, T.\,Kurth$^{a}$, 
L.\,Lellouch$^{e}$, T.\,Lippert$^{a,b}$, A. Portelli$^{e}$, A.\,Ramos$^{e}$,
A.\,Sch\"afer$^{d}$, K.K.\,Szab\'o$^{a}$
\end{center}

\vspace*{0.1cm}

\begin{flushleft}
\normalsize{$^a$Bergische Universit\"at Wuppertal, Gaussstr.\,20,
D-42119 Wuppertal, Germany}\\
\normalsize{$^b$J\"ulich Supercomputing Centre, Forschungszentrum J\"ulich,
D-52425 J\"ulich, Germany}\\
\normalsize{$^c$Institute for Theoretical Physics, E\"otv\"os University,
H-1117 Budapest, Hungary}\\
\normalsize{$^d$Universit\"at Regensburg, Universit\"atsstr.\,31,
D-93053 Regensburg, Germany}\\
\normalsize{$^e$Centre de Physique Th\'eorique\footnote{CPT is research unit
UMR 6207 of the CNRS and of Aix-Marseille univ. and Univ. Sud
Toulon-Var.}, Case 907, CNRS Luminy, 
F-13288 Marseille, France}
\end{flushleft}

\vspace*{0.1cm}

\begin{abstract}
  By using lattice QCD computations we determine the 
  sigma terms and strangeness content of all octet baryons by means of an
  application of the Hellmann-Feynman theorem. In addition to
  polynomial and rational expressions for the quark-mass dependence of
  octet members, we use $SU(3)$ covariant baryon chiral perturbation
  theory to perform the extrapolation to the physical up and down
  quark masses. Our $N_f=2+1$ lattice
  ensembles include pion masses down to about $190$ MeV in large
  volumes ($M_\pi L \gsim 4$), and three values of the lattice
  spacing.  Our main results are the nucleon sigma term
  $\sigma_{\pi N} = 39(4)(^{+18}_{-7})$ and the strangeness
  content $y_{N} = 0.20(7)(^{+13}_{-17})$.
  Under the assumption of validity of covariant baryon $\chi$PT in our
  range of masses one finds $y_{N} = 0.276(77)(^{+90}_{-62})$.
\end{abstract}

\end{titlepage}

%% file: stout.tex
\begin{tabular}{@{\,}l@{\,}@{\,}c@{\,}@{\,}c@{\,}@{\,}c@{\,}@{\,}c@{\,}@{\,}c@{\,}@{\,}c@{\,}@{\,}}
\hline
\hline
$\;\beta$ & $am_{ud}$ & $am_s$ & $L^3\!\times\!T$ &
Trajectories & $aM_\pi$ & $aM_K$ \\
\hline
\multirow{6}{*}{3.3}
 & -0.0960 & -0.057 & $16^3\!\times\!32$ & 10000 & 0.4115(6) & 0.4749(6) \\
 & -0.1100 & -0.057 & $16^3\!\times\!32$ &  1450 & 0.322(1) & 0.422(1)  \\
 & -0.1200 & -0.057 & $16^3\!\times\!64$ &  4500 & 0.2448(9) & 0.3826(6) \\
 & -0.1233 & -0.057 & $24^3\!\times\!64$ &  2000 & 0.2105(8) & 0.3668(6) \\
 & -0.1233 & -0.057 & $32^3\!\times\!64$ &  1300 & 0.211(1) & 0.3663(8) \\
 & -0.1265 & -0.057 & $24^3\!\times\!64$ &  2100 & 0.169(1) & 0.3500(7) \\
\hline
\multirow{4}{*}{3.57}
 & -0.0318 & 0,-0.010 & $24^3\!\times\!64$ & 1650,1650 & 0.2214(7),0.2178(5) & 0.2883(7),0.2657(5)  \\
 & -0.0380 & 0,-0.010 & $24^3\!\times\!64$ & 1350,1550 & 0.1837(7),0.1778(7) & 0.2720(6),0.2469(6) \\
 & -0.0440 & 0,-0.007 & $32^3\!\times\!64$ & 1000,1000 & 0.1348(7),0.1320(7) & 0.2531(6),0.2362(7) \\
 & -0.0483 & 0,-0.007 & $48^3\!\times\!64$ &  500,1000 & 0.0865(8),0.0811(5) & 0.2401(8),0.2210(5) \\
\hline
\multirow{5}{*}{3.7}
 & -0.007 & 0.0 & $32^3\!\times\!96$ & 1100 & 0.2130(4)  & 0.2275(4)   \\
 & -0.013 & 0.0 & $32^3\!\times\!96$ & 1450 & 0.1830(4)  & 0.2123(3)   \\
 & -0.020 & 0.0 & $32^3\!\times\!96$ & 2050 & 0.1399(3)  & 0.1920(3)   \\
 & -0.022 & 0.0 & $32^3\!\times\!96$ & 1350 & 0.1273(5)  & 0.1882(4)   \\
 & -0.025 & 0.0 & $40^3\!\times\!96$ & 1450 & 0.1021(4)  & 0.1788(4)   \\
\hline
\hline
\end{tabular}

%% file: gxa.tex
      \begin{tabular}{|c|c|c|c|c|}
        \hline 
         $\alpha$ &
        $g_{\alpha N}$ & $g_{\alpha \Lambda}$ & $g_{\alpha\Sigma}$ &
        $g_{\alpha \Xi}$ \\ 
        \hline
        \hline
         $\pi$ &
         $\frac{3}{4}g_A^2$ & $\frac{1}{(1+\xi)^2}g_A^2$ &
         $\frac{(1+6\xi^2)}{3(1+\xi)^2}g_A^2$ &
         $\frac{3(1-\xi)^2}{4(1+\xi)^2}g_A^2$ \\
        \hline
         $K$ &
         $\frac{(5-6\xi+9\xi^2)}{6(1+\xi)^2}g_A^2$ & 
         $\frac{(1+9\xi^2)}{3(1+\xi)^2}g_A^2$ &
         $\frac{(1+\xi^2)}{(1+\xi)^2}g_A^2$ &
         $\frac{(5+6\xi+9\xi^2)}{6(1+\xi)^2}g_A^2$ \\
        \hline
         $\eta$ &
         $\frac{(1-3\xi)^2}{12(1+\xi)^2}g_A^2$ & 
         $\frac{1}{3(1+\xi)^2}g_A^2$ &
         $\frac{1}{3(1+\xi)^2}g_A^2$ &
         $\frac{(1+3\xi)^2}{12(1+\xi)^2}g_A^2$ \\
        \hline
      \end{tabular}

%% file: lo.tex
      \begin{tabular}{|c|c|c|c|c|}
        \hline 
        $X$ & $N$ & $\Lambda$ & $\Sigma$ & $\Xi$ \\ 
        \hline
        \hline
         $c_X^\pi$ &
         $-2(2b_0+b_D+b_F)$ & $-4(b_0+b_D/3)$ &
         $-4(b_0+b_D)$ &
         $-2(2b_0+b_D-b_F)$ \\
        \hline
         $c_X^{\bar ss}$ &
         $-2(b_0 + b_D - b_F)$ & 
         $-2(b_0 + 4b_D/3)$ &
         $-2b_0$ &
         $-2(b_0+b_D+b_F)$ \\
        \hline
      \end{tabular}

%% file: res.tex
 \begin{tabular}{|l|c|c|c|c|c|}
  \hline
   & $\sigma_{\pi X}$ [MeV]& $\sigma_{\bar ss X}$ [MeV]&
  $y_X$ & $f_{ud X}$ & $f_{\bar ss X}$ \\
  \hline
  \hline
  $N$ & $39(4)(^{+18}_{-7})$
  & $67(27)(^{+55}_{-47}) $
  & $0.20(7)(^{+13}_{-17})$
  & $0.042(5)(^{+21}_{-8}) $
  & $0.072(29)(^{+60}_{-50}) $ \\
    \hline
  $\Lambda$ & $29(3)(^{+11}_{-5})$
  & $180(26)(^{+48}_{-77}) $
  & $0.51(15)(^{+48}_{-27})$
  & $0.027(3)(^{+10}_{-5}) $
  & $0.165(25)(^{+47}_{-63}) $ \\
  \hline
  $\Sigma$ & $23(3)(^{+19}_{-3})$
  & $245(29)(^{+50}_{-72}) $
  & $0.82(21)(^{+87}_{-39})$
  & $0.019(3)(^{+17}_{-3})$
  & $0.210(25)(^{+46}_{-62}) $ \\
  \hline
  $\Xi$ & $15(2)(^{+8}_{-3})$
  & $312(32)(^{+72}_{-77}) $
  & $1.7(5)(^{+1.9}_{-0.7})$
  & $0.012(2)(^{+6}_{-2})$
  & $0.240(26)(^{+58}_{-61}) $ \\
  \hline
\end{tabular}

%% file: piN.tex
  \begin{tabular}{lc}
    \hline
    \hline
    Source of systematic error & {error on $\sigma_{\pi N}$ [MeV]} \\
    \hline
    Chiral Extrapolation: &  \\
     - Pion mass range & $ 9.0 $ \\
     - Functional form & $ 5.5 $ \\
    Continuum extrapolation & $ 1.9 $ \\
    \hline
    \hline
  \end{tabular}

%% file: lec.tex
      \begin{tabular}{|c|c|c|c|c|c|}
        \hline 
         $M_0$ [GeV]& $b_0$ [GeV$^{-1}$]&
        $b_D$ [GeV$^{-1}$]& $b_F$ [GeV$^{-1}$]& $g_A$ & $\xi$ \\
        \hline
        \hline
         $0.75(15)$ &
         $-0.71(24)$ &
         $0.103(60)$ &
         $-0.359(72)$ &
         $0.92(13)$ &
         $0.402(90)$ \\
        \hline
      \end{tabular}

%% file: paper.bbl
\providecommand{\href}[2]{#2}\begingroup\raggedright\begin{thebibliography}{10}

\bibitem{Koch:1982pu}
R.~Koch, {\it {A New Determination of the pi N Sigma Term Using Hyperbolic
  Dispersion Relations in the (nu**2, t) Plane}},  {\em Z.Phys.} {\bf C15}
  (1982) 161--168.

\bibitem{Gasser:1990ce}
J.~Gasser, H.~Leutwyler, and M.~Sainio, {\it {Sigma term update}},  {\em
  Phys.Lett.} {\bf B253} (1991) 252--259.

\bibitem{Pavan:2001wz}
M.~Pavan, I.~Strakovsky, R.~Workman, and R.~Arndt, {\it {The Pion nucleon Sigma
  term is definitely large: Results from a G.W.U. analysis of pi nucleon
  scattering data}},  {\em PiN Newslett.} {\bf 16} (2002) 110--115,
  [\href{http://xxx.lanl.gov/abs/hep-ph/0111066}{{\tt hep-ph/0111066}}].

\bibitem{Ellis:2008hf}
J.~R. Ellis, K.~A. Olive, and C.~Savage, {\it {Hadronic Uncertainties in the
  Elastic Scattering of Supersymmetric Dark Matter}},  {\em Phys. Rev.} {\bf
  D77} (2008) 065026, [\href{http://xxx.lanl.gov/abs/0801.3656}{{\tt
  arXiv:0801.3656}}].

\bibitem{Giedt:2009mr}
J.~Giedt, A.~W. Thomas, and R.~D. Young, {\it {Dark matter, the CMSSM and
  lattice QCD}},  {\em Phys.Rev.Lett.} {\bf 103} (2009) 201802,
  [\href{http://xxx.lanl.gov/abs/0907.4177}{{\tt arXiv:0907.4177}}].

\bibitem{Borsanyi:2010bp}
{\bf Wuppertal-Budapest Collaboration} Collaboration, S.~Borsanyi {\em
  et.~al.}, {\it {Is there still any $T_c$ mystery in lattice QCD? Results with
  physical masses in the continuum limit III}},  {\em JHEP} {\bf 1009} (2010)
  073, [\href{http://xxx.lanl.gov/abs/1005.3508}{{\tt arXiv:1005.3508}}].

\bibitem{Durr:2010ni}
S.~Durr, Z.~Fodor, J.~Frison, T.~Hemmert, C.~Hoelbling, {\em et.~al.}, {\it
  {Sigma term and strangeness content of the nucleon}},  {\em PoS} {\bf
  LATTICE2010} (2010) 102, [\href{http://xxx.lanl.gov/abs/1012.1208}{{\tt
  arXiv:1012.1208}}].

\bibitem{Durr:2008rw}
S.~Durr {\em et.~al.}, {\it {Scaling study of dynamical smeared-link clover
  fermions}},  {\em Phys. Rev.} {\bf D79} (2009) 014501,
  [\href{http://xxx.lanl.gov/abs/0802.2706}{{\tt arXiv:0802.2706}}].

\bibitem{Durr:2008zz}
S.~Durr {\em et.~al.}, {\it {Ab Initio Determination of Light Hadron Masses}},
  {\em Science} {\bf 322} (2008) 1224--1227,
  [\href{http://xxx.lanl.gov/abs/0906.3599}{{\tt arXiv:0906.3599}}].

\bibitem{Lellouch:2009fg}
L.~Lellouch, {\it {Kaon physics: a lattice perspective}},
  \href{http://xxx.lanl.gov/abs/0902.4545}{{\tt arXiv:0902.4545}}.

\bibitem{Gasser:1983yg}
J.~Gasser and H.~Leutwyler, {\it {Chiral Perturbation Theory to One Loop}},
  {\em Annals Phys.} {\bf 158} (1984) 142.

\bibitem{Weinberg:1978kz}
S.~Weinberg, {\it {Phenomenological Lagrangians}},  {\em Physica} {\bf A96}
  (1979) 327. Festschrift honoring Julian Schwinger on his 60th birthday.

\bibitem{Gasser:1987rb}
J.~Gasser, M.~Sainio, and A.~Svarc, {\it {Nucleons with Chiral Loops}},  {\em
  Nucl.Phys.} {\bf B307} (1988) 779.

\bibitem{Jenkins:1990jv}
E.~E. Jenkins and A.~V. Manohar, {\it {Baryon chiral perturbation theory using
  a heavy fermion Lagrangian}},  {\em Phys.Lett.} {\bf B255} (1991) 558--562.

\bibitem{Borasoy:1996bx}
B.~Borasoy and U.-G. Meissner, {\it {Chiral expansion of baryon masses and
  sigma terms}},  {\em Annals Phys.} {\bf 254} (1997) 192--232,
  [\href{http://xxx.lanl.gov/abs/hep-ph/9607432}{{\tt hep-ph/9607432}}].

\bibitem{Donoghue:1998bs}
J.~F. Donoghue, B.~R. Holstein, and B.~Borasoy, {\it {SU(3) baryon chiral
  perturbation theory and long distance regularization}},  {\em Phys.Rev.} {\bf
  D59} (1999) 036002, [\href{http://xxx.lanl.gov/abs/hep-ph/9804281}{{\tt
  hep-ph/9804281}}].

\bibitem{Borasoy:1998uu}
B.~Borasoy, {\it {Sigma terms in heavy baryon chiral perturbation theory
  revisited}},  {\em Eur.Phys.J.} {\bf C8} (1999) 121--130,
  [\href{http://xxx.lanl.gov/abs/hep-ph/9807453}{{\tt hep-ph/9807453}}].

\bibitem{Young:2002ib}
R.~D. Young, D.~B. Leinweber, and A.~W. Thomas, {\it {Convergence of chiral
  effective field theory}},  {\em Prog.Part.Nucl.Phys.} {\bf 50} (2003)
  399--417, [\href{http://xxx.lanl.gov/abs/hep-lat/0212031}{{\tt
  hep-lat/0212031}}].

\bibitem{Dorati:2007bk}
M.~Dorati, T.~A. Gail, and T.~R. Hemmert, {\it {Chiral perturbation theory and
  the first moments of the generalized parton distributions in a nucleon}},
  {\em Nucl.Phys.} {\bf A798} (2008) 96--131,
  [\href{http://xxx.lanl.gov/abs/nucl-th/0703073}{{\tt nucl-th/0703073}}].

\bibitem{Bernard:2007zu}
V.~Bernard, {\it {Chiral Perturbation Theory and Baryon Properties}},  {\em
  Prog.Part.Nucl.Phys.} {\bf 60} (2008) 82--160,
  [\href{http://xxx.lanl.gov/abs/0706.0312}{{\tt arXiv:0706.0312}}].

\bibitem{Nakamura:2010zzi}
{\bf Particle Data Group} Collaboration, K.~Nakamura {\em et.~al.}, {\it
  {Review of particle physics}},  {\em J.Phys.G} {\bf G37} (2010) 075021.

\bibitem{Cheng:1997tt}
T.~Cheng and L.-F. Li, {\it {Chiral quark model of nucleon spin flavor
  structure with SU(3) and axial U(1) breakings}},  {\em Phys.Rev.} {\bf D57}
  (1998) 344--349, [\href{http://xxx.lanl.gov/abs/hep-ph/9701248}{{\tt
  hep-ph/9701248}}].

\bibitem{Capitani:2006ni}
S.~Capitani, S.~Durr, and C.~Hoelbling, {\it {Rationale for UV-filtered clover
  fermions}},  {\em JHEP} {\bf 0611} (2006) 028,
  [\href{http://xxx.lanl.gov/abs/hep-lat/0607006}{{\tt hep-lat/0607006}}].

\bibitem{Hoffmann:2007nm}
R.~Hoffmann, A.~Hasenfratz, and S.~Schaefer, {\it {Non-perturbative improvement
  of nHYP smeared Wilson fermions}},  {\em PoS} {\bf LAT2007} (2007) 104,
  [\href{http://xxx.lanl.gov/abs/0710.0471}{{\tt arXiv:0710.0471}}].

\bibitem{Luscher:1985dn}
M.~Luscher, {\it {Volume Dependence of the Energy Spectrum in Massive Quantum
  Field Theories. 1. Stable Particle States}},  {\em Commun.Math.Phys.} {\bf
  104} (1986) 177.

\bibitem{Leinweber:2003dg}
D.~B. Leinweber, A.~W. Thomas, and R.~D. Young, {\it {Physical nucleon
  properties from lattice QCD}},  {\em Phys.Rev.Lett.} {\bf 92} (2004) 242002,
  [\href{http://xxx.lanl.gov/abs/hep-lat/0302020}{{\tt hep-lat/0302020}}].

\bibitem{Procura:2003ig}
M.~Procura, T.~R. Hemmert, and W.~Weise, {\it {Nucleon mass, sigma term and
  lattice QCD}},  {\em Phys.Rev.} {\bf D69} (2004) 034505,
  [\href{http://xxx.lanl.gov/abs/hep-lat/0309020}{{\tt hep-lat/0309020}}].

\bibitem{Procura:2006bj}
M.~Procura, B.~Musch, T.~Wollenweber, T.~Hemmert, and W.~Weise, {\it {Nucleon
  mass: From lattice QCD to the chiral limit}},  {\em Phys.Rev.} {\bf D73}
  (2006) 114510, [\href{http://xxx.lanl.gov/abs/hep-lat/0603001}{{\tt
  hep-lat/0603001}}].

\bibitem{Ohki:2008ff}
H.~Ohki, H.~Fukaya, S.~Hashimoto, T.~Kaneko, H.~Matsufuru, {\em et.~al.}, {\it
  {Nucleon sigma term and strange quark content from lattice QCD with exact
  chiral symmetry}},  {\em Phys.Rev.} {\bf D78} (2008) 054502,
  [\href{http://xxx.lanl.gov/abs/0806.4744}{{\tt arXiv:0806.4744}}].

\bibitem{Young:2009zb}
R.~Young and A.~Thomas, {\it {Octet baryon masses and sigma terms from an SU(3)
  chiral extrapolation}},  {\em Phys.Rev.} {\bf D81} (2010) 014503,
  [\href{http://xxx.lanl.gov/abs/0901.3310}{{\tt arXiv:0901.3310}}].

\bibitem{MartinCamalich:2010fp}
J.~Martin~Camalich, L.~Geng, and M.~Vicente~Vacas, {\it {The lowest-lying
  baryon masses in covariant SU(3)-flavor chiral perturbation theory}},  {\em
  Phys.Rev.} {\bf D82} (2010) 074504,
  [\href{http://xxx.lanl.gov/abs/1003.1929}{{\tt arXiv:1003.1929}}].

\bibitem{WalkerLoud:2008bp}
A.~Walker-Loud, H.-W. Lin, D.~Richards, R.~Edwards, M.~Engelhardt, {\em
  et.~al.}, {\it {Light hadron spectroscopy using domain wall valence quarks on
  an Asqtad sea}},  {\em Phys.Rev.} {\bf D79} (2009) 054502,
  [\href{http://xxx.lanl.gov/abs/0806.4549}{{\tt arXiv:0806.4549}}].

\bibitem{Alexandrou:2008tn}
{\bf European Twisted Mass Collaboration} Collaboration, C.~Alexandrou {\em
  et.~al.}, {\it {Light baryon masses with dynamical twisted mass fermions}},
  {\em Phys.Rev.} {\bf D78} (2008) 014509,
  [\href{http://xxx.lanl.gov/abs/0803.3190}{{\tt arXiv:0803.3190}}].

\bibitem{Ishikawa:2009vc}
{\bf PACS-CS Collaboration} Collaboration, K.-I. Ishikawa {\em et.~al.}, {\it
  {SU(2) and SU(3) chiral perturbation theory analyses on baryon masses in 2+1
  flavor lattice QCD}},  {\em Phys.Rev.} {\bf D80} (2009) 054502,
  [\href{http://xxx.lanl.gov/abs/0905.0962}{{\tt arXiv:0905.0962}}].

\bibitem{Oller:2006yh}
J.~A. Oller, M.~Verbeni, and J.~Prades, {\it {Meson-baryon effective chiral
  lagrangians to O(q**3)}},  {\em JHEP} {\bf 0609} (2006) 079,
  [\href{http://xxx.lanl.gov/abs/hep-ph/0608204}{{\tt hep-ph/0608204}}].

\bibitem{Frink:2006hx}
M.~Frink and U.-G. Meissner, {\it {On the chiral effective meson-baryon
  Lagrangian at third order}},  {\em Eur.Phys.J.} {\bf A29} (2006) 255--260,
  [\href{http://xxx.lanl.gov/abs/hep-ph/0609256}{{\tt hep-ph/0609256}}].

\bibitem{Gasser:1984ux}
J.~Gasser and H.~Leutwyler, {\it {Low-Energy Expansion of Meson Form-Factors}},
   {\em Nucl. Phys.} {\bf B250} (1985) 517--538.

\bibitem{Andreas:LAT11}
A.~Sch\"afer, talk at Lattice 2011.

\bibitem{Roger:LAT11}
R.~Horsley, talk at Lattice 2011.

\end{thebibliography}\endgroup
